\documentclass[a4paper,12pt]{article}
\usepackage{amsmath, bm}
\usepackage{pstricks}
\usepackage{pst-node}
\usepackage[ansinew]{inputenc}
\usepackage{amssymb,amsmath}
\usepackage{amsfonts}
\usepackage{epsfig}

\usepackage{pst-node}

 \let\~\widetilde
\def\BSE{\begin{subequations}}\def\ESE{\end{subequations}}
\let \= \equiv

\def\p{\partial}
\def\px{\partial_x}
\def\py{\partial_y}
\def\a{\alpha}
\def\b{\beta}
\def\g{\gamma}
\def\d{\delta}
\def\o{\omega}
\def\wt{\widetilde}
\def\ms{\medskip}

\font\Sets=msbm10

\def\Integer {\hbox{\Sets Z}}    \def\Real {\hbox{\Sets R}}

\def\Complex {\hbox{\Sets C}}   \def\Natural {\hbox{\Sets N}}

\def\be{\begin{equation}}       \def\ba{\begin{array}}

\def\ee{\end{equation}}         \def\ea{\end{array}}

\def\bea {\begin{eqnarray}}      \def\eea {\end{eqnarray}}

\def\bean{\begin{eqnarray*}}    \def\eean{\end{eqnarray*}}

\def\pa  {\partial}             \def\ti  {\widetilde}

\def\la  {\lambda}

\def\eps{\varepsilon}           \def\ph{\varphi}

\def\const {\mathop{\rm const}\nolimits}

\def\tr    {\mathop{\rm trace}\nolimits}

\def\res   {\mathop{\rm res}  \limits}

\def\diag  {\mathop{\rm diag} \nolimits}

\def\im    {\mathop{\rm Im}   \nolimits}

\def\ker  {\mathop{\rm Ker} \nolimits}

\def\RA {\ \Rightarrow\ }         \def\LRA {\ \Leftrightarrow\ }

\def\qed   {\vrule height0.6em width0.3em depth0pt}

\def\<{\langle} \def\({\left(}  \def\>{\rangle} \def\){\right)}

\def\defeq {\stackrel{\mbox{\rm\small def}}{=}}

\newtheorem{exi}{Example}

\begin{document}

\author{E. Kartashova$^{a,\footnote{Corresponding author: lena@risc.uni-linz.ac.at}}$ ,  A. Kartashov$^b$  \\\\
$^a$ RISC, J.Kepler University, Linz, Austria\\
$^b$ AK-Soft, Linz, Austria }

\title{Laminated Wave Turbulence: Generic Algorithms I}

\date{}
 \maketitle


\newpage
Address:

Dr. Elena Kartashova

RISC, J. Kepler University

Altenbergerstr. 69

4020 Linz

Austria

e-mail: lena@risc.uni-linz.ac.at

Tel.: +42 (0)732 2468 9929

Fax: +42 (0)732 2468 9930

\newpage
\begin{abstract}
The model of laminated wave turbulence presented recently unites
both types of turbulent wave systems - statistical wave turbulence
(introduced by Kolmogorov and brought to the present form by
numerous works of Zakharov and his scientific school since nineteen
sixties) and discrete wave turbulence (developed in the works of
Kartashova in nineteen nineties). The main new feature described by
this model is the following: discrete effects do appear not only in
the long-wave part of the spectral domain (corresponding to small
wave numbers) but all through the spectra thus putting forth a novel
problem - construction of fast algorithms for computations in
integers of order $10^{12}$ and more. In this paper we present a
generic algorithm for polynomial dispersion functions and illustrate
it by application to gravity and planetary waves.

{\it Math. Classification:} 76F02,  65H10, 65Yxx, 68Q25

\end{abstract}

\maketitle

\def\p{\partial}
\def\px{\partial_x}
\def\py{\partial_y}
\def\a{\alpha}
\def\b{\beta}
\def\g{\gamma}
\def\d{\delta}
\def\o{\omega}
\def\wt{\widetilde}
\def\ms{\medskip}

\font\Sets=msbm10

\def\Integer {\hbox{\Sets Z}}    \def\Real {\hbox{\Sets R}}

\def\Complex {\hbox{\Sets C}}   \def\Natural {\hbox{\Sets N}}

\def\be{\begin{equation}}       \def\ba{\begin{array}}

\def\ee{\end{equation}}         \def\ea{\end{array}}

\def\bea {\begin{eqnarray}}      \def\eea {\end{eqnarray}}

\def\bean{\begin{eqnarray*}}    \def\eean{\end{eqnarray*}}

\def\pa  {\partial}             \def\ti  {\widetilde}

\def\la  {\lambda}

\def\eps{\varepsilon}           \def\ph{\varphi}

\def\const {\mathop{\rm const}\nolimits}

\def\tr    {\mathop{\rm trace}\nolimits}

\def\res   {\mathop{\rm res}  \limits}

\def\diag  {\mathop{\rm diag} \nolimits}

\def\im    {\mathop{\rm Im}   \nolimits}

\def\ker  {\mathop{\rm Ker} \nolimits}

\def\RA {\ \Rightarrow\ }         \def\LRA {\ \Leftrightarrow\ }

\def\qed   {\vrule height0.6em width0.3em depth0pt}

\def\<{\langle} \def\({\left(}  \def\>{\rangle} \def\){\right)}

\def\defeq {\stackrel{\mbox{\rm\small def}}{=}}

\newpage

\section{INTRODUCTION}
Statistical theory of wave turbulence begins with the pioneering
paper \cite{Kol1} of Kolmogorov presenting the energy spectrum of
turbulence  as a function of vortex size and thus founding the field
of mathematical analysis of turbulence. Kolmogorov regarded some
inertial range of wave numbers between viscosity and dissipation,
$k_0 < k < k_1$ for wave numbers $k$ where $k=|\vec{k}|$, and
suggested that in this range turbulence is locally homogeneous
 and  isotropic which, together with dimensional analysis,
allowed Kolmogorov to deduce that energy distribution is
proportional to $k^{-5/3}.$\\

Kolmogorov's ideas were further applied by Zakharov for construction
of wave kinetic equations \cite{lvov} which are approximately
equivalent to the initial nonlinear PDEs:
$$
\dot{A}_1=\int |V_{(123)}|^2
\delta(\o_1-\o_2-\o_3)\delta(\vec{k}_1-\vec{k}_2-\vec{k}_3)(A_2A_3-A_1A_2-A_1A_3)
{\bf d}\vec{k}_2 {\bf d}\vec{k}_3
$$
for 3-wave interactions, and similar equations for $i$-wave
interactions where $\delta$ is the Dirac delta-function and
$V_{(12..i)}$ is the vortex coefficient in the standard
representation of nonlinearity in the initial PDE: 
\be\label{sigma} \Sigma_{i} \frac{V_{(12..i)} \delta (\vec k_1 +\vec
k_2 + ... +\vec k_{i})} {\delta (\o_1+ \o_2+... + \o_i) } A_1A_2
\cdots A_i .\ee
The main idea of the wave turbulence theory is to take into account
only resonant interactions of waves described by
\begin{eqnarray}\label{open}
\begin{cases}
\omega (\vec k_1) \pm \omega (\vec k_2)\pm ... \pm \omega (\vec k_{n+1}) = 0,\\
\vec k_1 \pm \vec k_2 \pm ... \pm \vec k_{n+1} = 0
\end{cases}
\end{eqnarray}

Statistical wave turbulence theory deals with {\bf real solutions}
of Sys.(\ref{open}) and one of its  most important discoveries in
the statistical wave turbulence theory are  stationary exact
solutions of the kinetic equations first found in \cite{fil}. These
solutions have the form $k^{-\a}$ with $\a >0$  and are now called
Zakharov-Kolmogorov (ZK) energy spectra (see Fig.1). \\

The left part of Fig.1, the so-called finite length effects, have
been studied in the papers of Kartashova \cite{KarAll} where
properties of integer solutions of Sys.(\ref{open}) have been
studied. It was proven in particular that the spectral space of the
discrete wave system is decomposed into the small disjoint groups of
waves showing periodic energy fluctuations (nodes connected by lines
at the right panel of Fig. 2) and many waves with constant energy,
depicted by
blue diamonds.\\

The most important result of the theory of laminated wave turbulence
\cite{lam} is following: discrete effects do appear not only in the
long-wave part of the spectral space (corresponding to small wave
numbers) but all along the wave spectra. From the computational
point of view this theory gives rise to a completely novel problem:
 construction of fast algorithms for computations of integer
 solutions of Sys.(\ref{open})
in integers of order $10^12$ and more. For instance, for
 4-wave interactions of 2-dimensional gravity waves
this system has the form
$$ \sqrt{k_1}+\sqrt{k_2}=\sqrt{k_3}+\sqrt{k_4},\quad
\vec{k}_1+\vec{k}_2=\vec{k}_3+\vec{k}_4 ,$$ where
$\vec{k}_i=(m_i,n_i), \ \forall i=1,2,3,4$ and
$k_i=|\vec{k}_i|=\sqrt{m_i^2+n_i^2}$. It means that in a finite but
big enough domain of wave numbers, say $|m|,|n| \le D \sim 1000$,
direct approach leads to necessity to perform extensive
(computational complexity $D^8$) computations
with integers of the order of $10^{12}$.\\

Importance of the discrete layer of laminated turbulence is
emphasized by the fact that there exist many wave systems described
{\bf only} by discrete waves approach (for instance, wave systems
with periodic or zero boundary conditions). A sketch of the first
algorithm for
the problems of laminated turbulence is given in \cite{LamAlg}.\\

In this paper we present generic algorithm for computing discrete
layers of wave turbulent systems with dispersion function being a
function of the modulus of the wave vector $\vec{k}$, $\o=\o(k)$.
The main idea underlying our algorithm is the partition of the
spectral space into disjoint classes of vectors which allows us to
look for the solutions of Sys.(\ref{open}) in each class separately.
In Sec.2 we describe this construction in detail and with numerous
examples because its brief description given in \cite{AMS} is often
misunderstood by other researchers. In Sec. 3 the generic algorithm
is presented with gravity waves taken as our main example while in
Sec. 4 modification of this algorithm is given for the oceanic
planetary waves. Results of the computations and brief discussion
are given at the end.

\section{DEFINITION of CLASSES}

For a given $c \in \Natural, c \neq 0, 1, -1$ consider the set of
algebraic numbers $R_c= {±k^{1/c}, k \in \Natural}$. Any such number
$k_c$ has a unique representation
$$
    k_c = \g q^{1/c} , \g \in \Integer
$$
where $q$ is a product
$$
    q=p_1^{e_1} p_2^{e_2} ... p_n^{e_n},
$$
while $p_1, ... p_n$ are all different primes and the powers $e_1,
...e_n \in \Natural $ are all smaller than $c$.

\paragraph{Definition.}
    The set of numbers from $R_c$ having the same $q$ is called $q$-class $Cl_q$
    (also called "class $q$").
    The number $q$ is called class index. For a number $k_{(c)} = \g q^{1/c}$,
    $\g$ is called the weight of $k_{(c)}$.\\

The following two Lemmas are easily obtained from elementary
properties of algebraic numbers.

\paragraph{Lemma 1.}
    For any two numbers $k_1, k_2$ belonging to the same $q$-class,
    all their linear combinations
$$
    a_1 k_1 + a_2 k_2, \quad a_1, a_2 \in \Integer
$$
with integer coefficients belong to the same class $q$.\\

Indeed, if $k_1 = \g_1 q^{1/c}, k_2 = \g_2 q^{1/c}$ then
$$
    a_1 k_1 + a_2 k_2 =  \g_1 q_1^{1/c} + \g_2 q_2^{1/c} =
    (\g_1+\g_2)q^{1/c}
$$
in other words, every class is a one-dimensional module over the
ring of integers $\Integer$.

\paragraph{Example 1.}
Let us take $c = 2$.
    Numbers $k_1 = \sqrt{8}$ and $k_2 = \sqrt{18}$ belong to the same class
    $2$: $k_1 = 2\sqrt{2}, k_2 = 3\sqrt{2}$. According to Lemma 1, their sum $k_1 + k_2$
    belongs to the same class, and indeed, $k_1 + k_2 = \sqrt{50}=  5
    \sqrt{2}$.\\

\paragraph{Lemma 2.}
    For any $n$ numbers $k_1, k_2 ... k_n$ belonging to pairwise different
    $q$-classes, the equation
$$
k_1 \pm k_2 ...  \pm k_n = 0
$$
has no nontrivial solutions.\\

The statement of the Lemma 2 follows  from some known properties of
algebraic numbers and we are not going to present a detailed proof
here. The general idea of the proof is very simple indeed: a linear
combination of two different irrational numbers $\sqrt{q_1},
\sqrt{q_2}$ can not satisfy any equation with rational coefficients.
For example, equation $ a\sqrt{3}+b\sqrt{5}=0 $ has no solutions for
arbitrary rational $a$ and $b$. Indeed, taking the second power of
both sides we immediately come to a contradiction, as the irrational
number $\sqrt{15}$ is not equal to any rational number. As for the
general situation, one can apply the Besikovitch theorem \cite{bes}:

\paragraph{Theorem.}  Let
$$
a_1=b_1p_1, \  a_2=b_2p_2, \  ...,  \ a_s=b_sp_s,
$$
where $p_1, p_2, \ ..., \ p_s$ are different primes, and $b_1, b_2,
\ ...., \ b_s$ are positive integers not divisible by any of these
primes. If $x_1, x_2, \ ..., \ x_s$ are positive real roots of the
equations
$$
x^{n_1}- a_1=0, \ x^{n_2}- a_2=0, \ ... , \ x^{n_s}- a_s=0,
$$
and $P(x_1, \ x_2, \ ... , \ x_s)$ is a polynomial with rational
coefficients of degree less than or equal to $n_1-1$ with respect to
$x_1$, less that or equal to $n_2-1$ with respect to $x_2$, and so
on, then $P(x_1, \ x_2, \ ... , \ x_s)$ can vanish only if all its
coefficients vanish.

\paragraph{Corollary.}
    Any equation
$$
a_1 k_1 + a_2 k_2 ...  + a_n k_n = 0, a_i \in \Integer
$$
with $k_1, k_2, ... k_n$ belonging to pairwise different q-classes
has no nontrivial solutions.\\

Obviously, while $k_i$ can be represented as $\g_i q_i^{1/c}$, each
number $a_i k_i = sgn(a_i)|a_i| \g_i q_i^{1/c}$.

\paragraph{Example 2.}
Let us take $c = 2$.
    Numbers  $k_1 = \sqrt{8}$ and $k_2 = \sqrt{12}$
    belong to different classes: $k_1 = 2\sqrt{2}, k_2 = 2\sqrt{3}$ and according to
    Lemma 2 the equation $a_1 k_1 + a_2 k_2 = 0$ has only the trivial solution $a_1 = a_2 =
    0.$\\

The role of classes in the study of resonant wave interactions lies
in the following theorem.

\paragraph{Theorem 1.}
    Consider the equation
$$
a_1 k_1 + a_2 k_2 ...  + a_n k_n = 0, a_i \in \Integer
$$
where each $k_i = \g_i q_i^{1/c}$ belongs to some class $q_i \in
q_1, q_2 ... q_l, \quad l<n$. This equation is equivalent to a
system
\bea \label{sys2}
\begin{cases}a_{q_1,1} \g_{q_1,1} + a_{q_1,2}
\g_{q_1,2} + ...+ a
_{q_1,n_1} \g_{ q_1,n_1} = 0\\
a_{q_2,1} \g_{q_2,1} + a_{q_2,2} \g_{q_2,2} + ... + a _{q_2,n_2}
\g_{
q_2,n_2} = 0\\
...\\
a_{q_l,1} \g_{q_l,1} + a_{q_l,2} \g_{q_l,2} + ... + a _{q_l,nl} \g_{
q_l,nl} = 0
\end{cases}
\eea
\paragraph{Proof.}
    Let us re-write the Equation grouping numbers belonging to
    same class as follows:
$$
(k_{q_1,1} + k_{q_1,2} + ... + k_{q_1,n_1}) + (k_{q_2,1} + k_{q_2,2}
+ ... + k_{q_2,n_2}) + ... + (k_{q_l,1} + k_{q_l,2} + ... +
k_{q_l,n_l}) = 0
$$
so that all numbers in the $i$-th bracket belong to the same class
$q_i$ and all $q_i$ are different. From Lemma 1 it follows that each
sum $k_{q_i,1} + k_{q_i,2} + ... + k_{q_i,n_i}$is some number
$k_{q_i}$ of the same class $q_i$ and the equation can be re-written
as
$$
k_{q_1} + k_{q_2} + ... k_{q_l} = 0
$$
with all different $q_i$. It immediately follows from Lemma 2 that
$k_{q_1}=0, k_{q_2}=0 + ... k_{q_l}=0$.

\paragraph{Example 3.}   Now consider the equation
\be\label{ex3eq1} a_1 \sqrt{8} + a_2 \sqrt{12} + a_3 \sqrt{18} + a_4
\sqrt{24}  + a_5 \sqrt{48} = 0 \ee which is equivalent to \be
\label{ex3eq2}2a_1 \sqrt{2} + 2a_2 \sqrt{3} + 3a_3 \sqrt{2} + 2a_4
\sqrt{6}  + 4a_5 \sqrt{3} = 0 \ee and according to the Theorem 1 is
equivalent to the system
\bea \label{ex3sys1} \begin{cases} 2 a_1 + 3 a_3 = 0\\
 2 a_2 + 4 a_5 = 0\\
  a_4 = 0
\end{cases}
\eea 
which is in every respect much simpler than the original equation.\\

    The computational aspect of this transformation is an especially important
    illustration to the main idea of the algorithm presented in this paper.
    Suppose we are to find all {\bf exact} solutions of
    (\ref{ex3eq1})
    in some finite domain $1 \le a_i \le D$. Then the straightforward iteration
    algorithm needs  $O(D^4)$ floating-point operations, $a_i=1..D, i=1,2,3,4$
    (even ignoring difficulties with floating-point arithmetic precision for large $D$).
    On the other hand, solutions of Sys.(\ref{ex3sys1}) can, evidently,
    be found in $O(D)$ operations with integer numbers.

\section{EXAMPLE ONE: GRAVITY WAVES}

    To show the power of the approach outlined above in practice,
    we proceed as follows. First we give a detailed description of the
    algorithm which is used to find all physically relevant four-wave interactions
    of the so-called gravity waves in a finite domain. We also
    estimate computational complexity and memory requirements for
    its implementation and present results of our computer
    simulations.
    In the next section we discuss reusability of this algorithm and
    transform it to solve a similar problem for three-wave
    interactions of planetary waves.
    Further we briefly discuss applicability of our algorithm to
    other wave-type interactions.

\subsection{Problem Setting}
    The main object of our studies are four-tuples of {\it gravity
    waves}. A wave is characterized by its {\it  wave vector}
    $\vec{k}$, a two-dimensional vector with integer coordinates
    $\vec{k}=(m,n), \quad m,n \in
    \Integer$, where $m,n$ may not both be zero. We define the usual
    norm of this vector $k=|\vec{k}|=\sqrt{m^2+n^2}$ and its {\it
    dispersion function} $w_k=\sqrt{k}$.

\paragraph{Definition.} Four wave vectors
    $\vec{k}_1, \vec{k}_2, \vec{k}_3, \vec{k}_4$ are called a
    {\it resonantly interacting four-tuple} if the following
    conditions are fulfilled:
\bea\label{prosetdef1} \begin{cases}
\sqrt{k_1}+\sqrt{k_2}=\sqrt{k_3}+\sqrt{k_4} \\
\vec{k}_1+\vec{k}_2=\vec{k}_3+\vec{k}_4
\end{cases}\eea
where $\vec{k}_i=(m_i,n_i), \ \forall i=1,2,3,4$ and
$k_i=|\vec{k}_i|=\sqrt{m_i^2+n_i^2}$.\\

Sys. (\ref{prosetdef1}) is written in vector form and is equivalent
to the following system of scalar equations:
\bea\label{prosetdef2} \begin{cases}
{(m_1^2+n_1^2)}^{1/4} + {(m_2^2+n_2^2)}^{1/4}={(m_3^2+n_3^2)}^{1/4}+{(m_4^2+n_4^2)}^{1/4} \\
m_1+m_2=m_3+m_4\\
n_1+n_2=n_3+n_4\\
\end{cases}\eea
Sometimes (especially in numerical examples) it is convenient to
represent the solution four-tuple as
\be\label{repfourw}
(m_{1L},n_{1L})(m_{2L},m_{2L})=(m_{1R},n_{1R})(m_{2R},m_{2R}) \ee
We are going to find all resonantly interacting four-tuples with
coordinates $m_i,n_i$ such that $-D \le m_i, n_i \le D, \quad
i=1,2,3,4$ for some $D \in \Natural$. The set of numbers $d \in
[-D,D]$ is further called  the main domain or simply domain.

\subsection{Computational Preliminaries}
\subsubsection{Strategy Choice}
Numerically solving irrational
equations in whole numbers is always
an intricate business. Basically, two approaches are widely used.\\

The first approach is to get rid of irrationalities (for equations
in radicals typically taking the expression to a higher power,
re-grouping members etc.). For an equation like $a\sqrt{x}
=b\sqrt{y}$ this approach is reasonable: we simply raise both sides
to power 2 and solve the equation $a^2x=b^2y$. (Some attention
should be payed to the signs of $a,b$ afterwards.) However, for
Sys.(\ref{prosetdef2}), containing four fourth-degree roots, this
approach is out of question.\\

The second approach is, to solve the equations using floating-point
arithmetic, obtain (unavoidably) approximate solutions  and develop
some (domain dependent) lower estimate for the deviation, which
would enable us to sort out exact solutions with deviation due only
to the floating-point. As an example, consider the equation
$\sqrt{x} = y$ in the domain $0 \le x,y \le D$. If $x$ is {\it not}
a square then $|\sqrt{x}-[\sqrt{x}]| \ge 1/{2\sqrt(1/D)} - 1/{8D}$,
so each solution with smaller deviation is a perfect square. In
other words, very small deviations are guaranteed to be an artefact
of floating point arithmetic.\\

This approach is more reasonable, though for Sys.(\ref{prosetdef2})
the corresponding estimate would probably be not so easy to obtain.
However, it has one crucial drawback, namely, its high computational
complexity. Indeed, Sys.(\ref{prosetdef2}) consists of 3 equations
in 8 variables and exhaustive search takes at least $O(D^5)$
operations, and many time
consuming operations (like taking fractional powers) at that.\\

Our primary goal is to find all solutions in the presently
physically relevant domain $D \widetilde 10^3$ with a possibility of
extension to larger domains. The algorithm should be {\it generic},
i.e. applicable to a wide class of wave types by simple
transformations. Studying resonant interactions of other physically
important waves we may have to deal with even more variables, e.g.
for inner waves in laminated fluid $\vec{k}=(m,n,l)$ and for
four-wave resonant interactions the brute force algorithm described
above has
computational complexity $O(D^8)$.\\

Clearly we need a crucially new algorithm to cope with the
situation; and here classes come to our aid.

\subsubsection{Application of Classes} Theorem 1, applied to the
first equation of  Sys.(\ref{prosetdef2})
 readily yields the following result.

\paragraph{Theorem 2.}
Given an equation
\be\label{th2eq1}
\sqrt[4]{t_1}+\sqrt[4]{t_2}=\sqrt[4]{t_3}+\sqrt[4]{t_4} \ee
with $t_i \in \Natural, t_i > 0$, two situations are possible:

\paragraph{Case 1:} all the numbers $t_i, \quad i=1,2,3,4$ belong
to the same class $Cl_{q}$.\\

In this case Eq.(\ref{th2eq1}) cab be rewritten as
\be\label{th2eq2} \g_1\sqrt[4]{q} +
\g_2\sqrt[4]{q}=\g_3\sqrt[4]{q}+\g_4\sqrt[4]{q} \ee
with $\g_1, \g_2, \g_3, \g_4 \in \Natural$ and $q$ an $R_4$ class
index (i.e. a natural number not divisible by a fourth degree of any
prime).

\paragraph{Case 2:} the vectors belong to two different classes
$Cl_{q_1},Cl_{q_2}$.\\

In this case Eq.(\ref{th2eq1}) can be rewritten as
\be\label{th2eq3} \g_1\sqrt[4]{q_1} +
\g_2\sqrt[4]{q_2}=\g_1\sqrt[4]{q_1}+\g_2\sqrt[4]{q_2} \ee
with $\g_1, \g_2 \in \Natural$ and $q_1, q_2$ being $R_4$ class
indexes.\\

Now we concentrate on Case 1 being most interesting physically. For
this case we can do computations class-by-class, i.e. for every
relevant $q$ we take all solutions of $\g_1+\g_2=\g_3+\g_4$ such
that $\g_i^4q$ can be represented as a sum of squares $\g_i^4q =
m_i^2+n_i^2, \quad |m_i|, |n_i| \le D$ and for every decomposition
into such sum of squares we check the linear condition
(\ref{prosetdef2}.2).

\subsection{Algorithm Description}
 At the beginning we have to
compute a very important domain-dependent parameter we need for the
computations.

Notice that in the main domain $-D \le m,n \le D$ every number under
the radical $t_i \le 2D^2$ i.e, $\g_i^4q \le 2D^2$. For a given $q$,
$\g^{max}(q) \le (2D^2/q)^{1/4}$.

\paragraph{Definition.}
A number  $\g^{max}(q)$ is called {\it class multiplicity} and
denoted $Mul(q)$. \\

For the main domain $D=10^3$, class multiplicities are reasonably
small numbers, $\g^{max}(1)=37$ being the largest. Class
multiplicities for the majority of classes (starting with
$q=125002$) are equal to $1$ - this fact will be later used to
achieve a major shortcut in computation time.

\subsubsection{Step 1. Calculating Relevant Class Indexes} Class
indexes of the module $R_c$ as defined above are numbers not
divisible by any prime in $c$-th degree, in our case $c=4$ not
divisible by  4-th power of any prime. We can further restrict
relevant class indexes as follows.\\

First, in (\ref{th2eq1}) every number under the radical
$t_i=\g_i^4q$ must have a representation as a sum  of two squares of
integer numbers, $t_i=m_i^2+n_i^2$. According to the well-known
Euler's theorem an integer can be represented as a sum of two
squares if and only if its prime factorization contains every prime
factor $p\equiv 4u+3$ in an even degree. As $\g_i^4$ evidently
contains every prime factor in an even degree, this condition must
also hold for $q$. This can be formulated as follows: if $q$ is
divisible by a prime $p\equiv 4u+3$, it should be divisible by its
square and should {\it
not} be divisible by its cube.\\

The implementation of this step is accomplished with a sieve-type
procedure. Create an array $Ar_q=[1,...2D^2]$ of binary numbers,
setting the all the elements of the array to $1$. Make the first
pass: for all primes $p$ in the region $2 \le p \le \sqrt[4]{2D^2}$
set to 0 the elements of the array $p^4, 2p^4, ... \kappa p^4$ where
$\kappa=\lfloor 2D^2/p^4\rfloor$. In the second pass, for all primes
$p_{4u+3} \equiv 3 \mod 4, p \le 2D^2$ and integer factors
$a=1...a_{max}$ such that $ap \le 2D^2$, do the following. If $a
\neq 0 \mod p$ then set the $ap$-th element of the array $Ar_q$ to
$0$. If $a \equiv 0 \mod p$, then if $a \equiv 0 \mod p^2$ then also
set the $ap$-th element of the array $Ar_q$ to $0$.\\

Notice that in the second pass the first check should only be done
for primes $p \le \sqrt{2D^2}$ and the second one - for $p \le
\sqrt[3]{2D^2}$.\\

We create an array $W_q$ of "work indexes". In the third pass, we
fill it with indexes of the array $Ar_q$ for which the elements'
values have not been set to 0 in the first two passes. We also
create an array of class multiplicities $Mul_q$ and fill it with
corresponding class multiplicities (see previous subsection).

\paragraph{Remark 1.} All numbers $q$ found above do have a
representation as a sum of two squares; however, some do not have
representation with $|m| \le D$ {\it and} $|n| \le D$. However, we
do not look for them now: they will be discarded automatically at
further steps.\\

The computational complexity of this step can be estimated in the
following way. The number of primes $\le x$ is, asymptotically,
$\pi(x) = x/\log(x)$, so their density around $x$ is $1/\log(x)$.
The first pass takes $\lfloor2D^2/p^4\rfloor$ operations for each
prime $2 \le p \le \sqrt[4]{2D^2}$ so the overall number of
operations can be estimated as
$$
\int_2^{\sqrt[4]{2D^2}}\frac{2D^2/x^4}{\log(x)}dx = O(D^2)
$$
As for the second pass, primes $p \equiv 4u+3 < 2D^2$ constitute
about a half of all primes and are evenly distributed among them.
Sieving out by a prime $p$ requires $O(2D^2/p) + O(2D^2/p^2) +
O(2D^2/p^3) = O(2D^2/p)$ operations which again boils down to
overall $O(D^2)$
steps.\\

Evidently, the third pass requires the same $O(D^2)$ operations.\\

So the overall computational complexity of this step is $O(D^2)$.

\paragraph{Remark 2.} It is not so easy to give a good estimate for the
number of class indices $\pi_{cl}(D)$. Of course $O(D^2/\ln(D) \le
\pi_{cl}(D) \le O(D^2)$ holds, and most probably  $\pi_{cl}(D) =
O(D^2/\ln(D)$. (This is presently under study.) Whenever we need
this number for estimating computational complexity of the
algorithm, we presume $\pi_{cl}(D) =O(D^2)$ to be on the safe side
of things. In our main computation domain $D=10^3$ the number of
class indices $\pi_{cl}(10^3) = 384145$.

\subsubsection{Step 2. Finding Decompositions into Sum of Two
Squares}

In 1908, G. Cornacchia \cite{corn} proposed an algorithm for solving
the diophantine equation $x^2+dy^2=4p$ with $p$ prime, $p=4u+1$.
This has been recently generalized to solving $x^2+dy^2=m$, $m$ not
necessarily prime \cite{basil}. To find all decompositions of a
number $\g^4q$ into two squares we can use a simplified variant of
this setting $d=1$. A very efficient implementation of this
algorithm can be obtained thanks to the following result
\cite{basil}:

\paragraph{Theorem 3.}
Let $t^2 \equiv -1 (\mod m), 0 < t < (m/2)$. Set $r_0=m$ and $r_1=t$
and construct the finite sequence $\{r_i\}, \quad
r_i=q_ir_{i+1}+r{i+2}, q_i=\lfloor r_i/r_{i+1}\rfloor$, for $0 \le i
\le n-1$, where $r_0 > r_1 > ... > r_n = 1 > r_{n+1} = 0$.
If $r^2_{k-1} > m > r^2_k$ then $m=r^2_k +r^2_{k+1}$.\\

The proof is based on elementary number-theoretic considerations.
Now it is evident that for each $0 < t < (m/2)$ such that $t^2
\equiv -1 (\mod m)$ we obtain one decomposition of $m$ into two
squares and the algorithm gives all decompositions with $x>y$. For
our use, we also take symmetrical decompositions $x<y$ and also
$x=y$ if $m=2x^2$. The computational complexity $T$ of the algorithm
is, basically, the complexity of finding all square roots of $-1$
modulo $m$ and is logarithmic in $m$, i.e. $T=O(\log(m))$.\\

Let $Dec_q$ be the maximal number of decompositions of $\g^4q, \quad
\g=1...Mul(q)$ into sum of two squares. We create a
three-dimensional array $Ar_D[G_q=1..Mul(q), D_q=1..Dec_q, 2]$ and
for each $G_q$ store the list of decompositions $(m_{G_q, D_q},
n_{G_q, D_q})$. We also create a one-dimensional auxiliary array
$Ar_{DtoW}$ storing the number of two-square decompositions for each
weight. The number of decompositions of an integer into sum of two
squares can be estimated as $O(\log(m))$ using the following Euler
theorem:

\paragraph{Theorem.} Let $m$ be a positive integer, and let
$$
m = 2^rp_1^{s1}...p_k^{sk}q_1^{t1}...q_l^{tl}
$$
be its factorization into prime numbers, where $p_i \equiv 1 (\mod
4)$ and $q_i \equiv 3 (\mod 4)$. Then the number of essentially
different decompositions of $m$ into sum of $2$ squares is equal to
the integral part of $\delta/2$ where
$$
\delta = (\prod_{j=1}^{k}(s_i+1))(\prod\frac{(-1)^{t_j}+1}{2}).
$$
Now we see that filling the array $Ar_D$ can be accomplished in
\be\label{time_sq} T=O(\log(q)+ \log(2^4q)+...+\log(Mul(q)^4q))
\ee
steps.\\

Using presentation (\cite{introw}, Eq.(4.4.8.1))
$$
\sum_{k=1}^n \ln(ak+b)= n\ln a + \ln\Gamma(b/a+n+1)-\ln\Gamma(b/a+1)
$$
 and the well-known
$$
n! \sim \sqrt{\pi n}(\frac{n}{e})^n
$$
we obtain $T=O(Mul(q)(\log q+\log(Mul(q)))=$$O(Mul(q)\log q)$. This
is much less than $O(Mul(q)^3)$ (see the next subsection) and
contribution of this step into the overall computational complexity
of the algorithm is negligible.

\subsubsection{Step 3. Solving the Sum-of-Weights Equation}

Consider now the equation for the weights
\be\label{4weights0} \g_1+\g_2=\g_3+\g_4 \ee
with $1 \le \g_i \le Mul(q)$ (see \ref{th2eq2}). For convenience we
change our notation to $\g_{1L}, \g_{2L}$ (left) and $\g_{1R},
\g_{2R}$ (right) and introduce weight sum $S_\g=\g_{1*}+\g_{2*}$.
Without loss of generality we can suppose
\be\label{4weights} \g_{1L} \le \g_{1R} \le \g_{2R} \le \g_{2L}.
\ee
Notice that we may not assume strict inequalities because even for
$\g_i=\g_j$ there may exist two distinct vectors $(m_i, n_i),
(m_j,n_j)$ with $m_i^2+n_i^2= m_j^2+n_j^2=\g^4q$ either due to the
possibility of representing $\g^4q$ as sum of two squares in
multiple ways or even for a single two-square representation - to
the possibility of taking different sign combinations $(\pm |m|, \pm
|n|)$ left and right.\\

Now we may encounter the following four situations (see Fig.1 a-d):
\begin{enumerate}
  \item {$\g_{1L} < \g_{1R} < \g_{2R} < \g_{2L} \quad$} (Fig.\ref{f:1a})\\
The general, physically most interesting case. Every solution yields
four waves with pairwise distinct modes.
  \item {$\g_{1L} = \g_{1R} < \g_{2R} = \g_{2L}\quad$} (Fig.\ref{f:1b})\\
  \item {$\g_{1L} < \g_{1R} = \g_{2R} < \g_{2L}\quad$} (Fig.\ref{f:1c})\\
  \item {$\g_{1L} = \g_{1R} = \g_{2R} = \g_{2L}\quad$} (Fig.\ref{f:1d})\\
  The "most degenerate" case.
\end{enumerate}

The search is organized as follows. Each admissible sum of weights
$S_\g; \quad 2 \le S_\g \le Mul(q)$ is partitioned into sum of two
numbers $S_\g = \g_{1L} + \g_{2L}, \quad 1 \le \g_{1L} \le \g_{2L}
\le Mul(q)$. Then the same number is partitioned into sum of
$\g_{1R}, \g_{2R}, \quad \g_{1L} \le \g_{1R} \le \g_{2R}  \le
\g_{2L} $. Evidently, if $S_{\g} \le Mul(q)+1$ then the minimal
$\g_{1L}$ is $1$, otherwise it is $S_{\g} - Mul(q)$ (to provide
$\g_{1R} \le Mul(q)$). The maximal $\g_{1L}$ is always $\lfloor
S_\g/2 \rfloor$ and
similarly $\g_{1R} \le \lfloor S_\g/2 \rfloor$.\\

The computational complexity of this step can be estimated as $T =
O(Mul(q)^3)$ due to $O(Mul(q))$ possibilities for each of three
values $S_\g, \g_{1L}, \g_{1R}$. As $Mul(q) =
\lfloor(2D^2/q)^{1/4}\rfloor$, $\quad T=O((2D^2/q)^{3/4})$.

\paragraph{Remark.} This step contains an evident redundancy. Indeed, the
equation \ref{4weights0} need not be solved independently for each
class. Instead, its solutions for all $S_\g$ could be computed in
advance and stored in a look-up table. However, this involves
significant computational overhead (e.g. the lookup procedure
includes computing the minimal $\g_{1L} = max(1, S_{\g} - Mul(q))$,
which {\bf must} be done for each class) wiping out the gains of
this approach, at least for our basic domain $D=10^3$. Nevertheless,
this approach should be kept in view if need for computations in
much larger domains, say $D=10^6$, arises.

\paragraph{Remark.} The general case is not really so general - most
classes have small multiplicities and then degenerate cases prevail.
The overall distribution is given below:\\

\begin{tabular}{|c|c|c|c|c|}
\hline 
    $Case$ & $1$ & $2$ & $3$ & $4$\\
\hline
    $Classes$ & 24368 & 57666 & 13987  & 63778\\
\hline 
\end{tabular}\\

\subsubsection{Step 4. Discarding "Lean" Classes}
In the main domain $D \le 10^3$ we encounter 384145 classes. This
sounds like a lot - however, most of these can be processed without
computations or with very simple computations. Notice the simple
fact that if a class has multiplicity $1$, Sys.(\ref{prosetdef2})
takes the form
\bea\label{prosetdeflean} \begin{cases}
q = m_{1L}^2+n_{1L}^2 = m_{2L}^2+n_{2L}^2 = m_{1R}^2+n_{1R}^2 = m_{2R}^2+n_{2R}^2 \\
m_{1L}+m_{2L}=m_{1R}+m_{2R}\\
n_{1L}+n_{2L}=n_{1R}+n_{2R}\\
\end{cases}\eea
and for any nontrivial solution the four vectors $(m_i, n_i)$ should
be pairwise distinct. In terms of the weight equation of the
previous section it means that solutions, if any, have to belong to
the fourth ("most degenerate") case. It is evident that no solution
of Sys.(\ref{prosetdeflean}) with pairwise distinct $(m_i, n_i)$
exist for $q$ having few decompositions into sum of two squares: one
($q = m^2+m^2$), two ($q = m^2+n^2 = n^2+m^2$) and three ($q =
m^2+n^2 = n^2+m^2 = l^2 + l^2$). It can be shown by means of
elementary algebra that this also holds for $q$ having four
decompositions.

\paragraph{Remark.} It is very probable that for classes of multiplicity 1
no nontrivial solutions exist, whatever the number of decompositions
into sum of two squares. The question is presently under study. In
the main domain $D=10^3$ we encounter 357183 classes of multiplicity
$1$ (1-classes). This is about $93\%$ of all classes in the domain.
Among them, the number of decompositions into sum of two squares is
distributed as follows:\\

\begin{tabular}{|c|c|c|c|c|c|c|c|c|c|}
\hline 
    $Dec(q)$ & $0$ & $1$ & $2$ & $3$ & $4$ & $5$ & $6$ & $7$ & $8$\\
\hline
    $Classes$ & 110562 & 256 & 138044  & 163  & 78886 & 3 & 8727 & 2 & 16595\\
\hline 
\end{tabular}\\

\begin{tabular}{|c|c|c|c|c|c|c|c|c|c|c|}
\hline 
    $Dec(q)$ & $14$ & $16$ & $18$ & $20$ & $24$ & $26$ & $32$ & $9$ & $10$ & $12$\\
\hline
    $Classes$ & 38 & 1015 & 84  & 1  & 75 & 1 & 1 & 31  & 269 & 2429\\
\hline 
\end{tabular}\\

{\small Table 1. Distribution of decomposition number $Mul(q)$ for
1-classes $q$ in the main domain $D=1000$}\\

It follows that 327911 1-classes can be discarded without any
computations at all and only 29272 must be checked for probable
solutions.

\subsubsection{Step 5. Checking Linear Conditions: Symmetries and
Signs}

Sum-of-weights equation solved and decompositions into sum of two
squares found, we need only check the linear conditions to find all
solutions. On the face of it, the step is trivial, however some
underwater obstacles have to be taken into account. Having found a
four-tuple of vectors $(m_i, n_i)$ satisfying the first equation of
Sys.(\ref{prosetdef2}) with both coordinates {\it non-negative},
solutions of the system will be found taking all combinations of
signs satisfying
\bea\label{linsigns} \begin{cases}
\pm m_1 \pm m_2= \pm m_3 \pm m_4\\
\pm n_1 \pm n_2= \pm n_3 \pm n_4\\
\end{cases}\eea
Even the straightforward approach does not need more than $2^8$
comparisons, so this step does not consume very much computing time.
However, a few points may not be overlooked in order to organize
correct, exhaustive and efficient search:

\begin{itemize}

\item{} For the system to represent a four-wave interaction, all the
four waves must be pairwise unequal. For the first degenerate case
of the weight diagram (Fig.\ref{f:1b}) we must provide $m_{1L} \neq
m_{1R}$ and $m_{2R} \neq m_{2R}$. For the second one
(Fig.\ref{f:1c}) - $m_{1R} \neq m_{2R}$ and for total degeneration
(Fig.\ref{f:1d}) - $m_{1L} \neq m_{2L} \neq m_{1R} \neq m_{2R}$.

\item{}One and the same solution may not occur among the 256 sign
combinations twice. First, this could happen due to some $m_i$ or
$n_i$ being $0$ (evidently, the $\pm$ variation should not be done
for any $0$ coordinate). Next, sign variation could lead to a
transposition of wave vectors. For example, for $q=1$ and
$\g_1+\g_2=10$ we obtain solutions
\bea \begin{cases}
(0,-9)(0,49) \RA (15,20)(-15,20) \nonumber \\
\mbox{and}\\
(0,-9)(0,49) \RA (-15,20)(15,20) \nonumber \\
\end{cases} \eea

which really represent one and the same four-tuple.

\item{} The set of solutions possesses some evident symmetries: if
\be\label{solpos} (m_{1L},n_{1L})(m_{2L},m_{2L}) \RA
(m_{1R},n_{1R})(m_{2R},m_{2R}) \nonumber \ee
then, of course,
 \be\label{solpos}
(-m_{1L},n_{1L})(-m_{2L},n_{2L})  \RA
(-m_{1R},n_{1R})(-m_{2R},n_{2R}) \nonumber \ee
\be\label{solpos} (m_{1L},-n_{1L})(m_{2L},-n_{2L}) \RA
(m_{1R},-n_{1R})(m_{2R},-n_{2R}) \nonumber \ee
\be\label{solpos} (-m_{1L},-n_{1L})(-m_{2L},-n_{2L}) \RA
(-m_{1R},-n_{1R})(-m_{2R},-n_{2R}) \nonumber \ee
\end{itemize}

Taking into account these points, an effective search is constructed
easily.

\section{EXAMPLE TWO: PLANETARY WAVES}

In this Section we demonstrate the flexibility of our algorithm.
Namely, we solve essentially the same problem - finding all integer
solutions in a finite domain $-D \le m,n \le D$ for three-wave
interactions and another wave type (the so-called "planetary
waves"). In this case $c=-2$ and the main equation, corresponding to
(\ref{prosetdef2}), has the form
\bea\label{pwdef1} \begin{cases}
\frac{1}{\sqrt{m_1^2+n_1^2}}+\frac{1}{\sqrt{m_2^2+n_2^2}}=\frac{1}{\sqrt{m_3^2+n_3^2}} \\
m_1 + m_2=m_3
\end{cases}\eea
where $|m_i|, |n_i| \le D$.

\subsection{Steps that Stay}
\begin{itemize}
\item{ Step 1 } - sieving out possible class bases -
undergoes minimal changes. For $c=-2$, each $q$
should be a square-free number and not divisible by any prime
$p=4u+3$. Evidently, for this wave type the set of class indices is
a subset of class indices of the previous section.

\item{ Step 2 } - decomposition into two squares -
can be preserved one-to-one. Indeed, there are sophisticated
algorithms for representing square-free numbers as sums of two
squares that are slightly more efficient than in the general case
(one used in the previous section) but this step is not the
bottleneck of the algorithm.
\end{itemize}

\subsection{Steps to be Modified}
\begin{itemize}
\item{ Step 3 } - the weight equation is in this case
\be\label{weights3r} \frac{1}{\g_1}+\frac{1}{\g_2}=\frac{1}{\g_3}
\ee
or
\be \g_3 = \frac{\g_1\g_2}{\g_1+ \g_2} \ee
which has relatively few solutions in integers. Indeed, even for
class $1$ with multiplicity $1414$ we obtain only $3945$
solutions.

\paragraph{Remark.} For this example it makes sense to generate and store
the set of triads $(\g_1, \g_2, \g_3)$ which constitute integer
solutions of the Eq.(\ref{weights3r}) for $1 \le \g_i \le Mul(1)$
and for each class $q$ just take its subset $1 \le \g_i \le Mul(q)$.

\item{ Step 4 } - discarding "lean" classes - becomes trivial: no
class with multiplicity $1$ yields an integer solution of
\ref{weights3r}. We need only consider $63828$ classes
($q_{63828}=499993, \quad q_{63829}=500009 $) from $243143$ in the
main domain.

\item{ Step 5 } - checking linear conditions - is also much easier
than in the previous example, i.e. one equation with three variables
instead of two equations with four variables each.
\end{itemize}

\section{DISCUSSION}
Our algorithm has been implemented in VBA programming language;
computation time (without disk output of solutions found) on a
low-end PC (800 MHz Pentium III, 512 MB RAM) is about 4.5 minutes
for Example 1 and 1.5 minutes for Example 2. Some overall numerical
data for both examples is given in the Tables and Figures below:\\

\begin{tabular}{|c|c|c|c|c|c|}
\hline 
    $Domain$ & $\le 200$ & $\le 400$ & $\le 600$ & $\le 800$ & $\le 1000$\\
\hline
    $Solutions$ & 263648 & 800435 & 932475  & 1127375  & 1389657\\
\hline 
\end{tabular}\\

{\small Table 2. {\bf Example 1:} Distribution of the number of
solutions depending on the chosen main domain $D$.}\\

It is interesting that though the overall number of solutions grows
sublinearly as we extend the domain, the number of asymmetrical
solutions ($\g_1 \ne \g_2 \ne \g_3 \ne \g_4$),
physically most important ones, grows faster than linearly:\\

\begin{tabular}{|c|c|c|c|c|c|}
\hline 
    $Domain$ & $\le 200$ & $\le 400$ & $\le 600$ & $\le 800$ & $\le 1000$\\
\hline
    $Solutions$ & 96 & 344 & 744  & 1328  & 2088\\
\hline 
\end{tabular}\\

{\small Table 3. {\bf Example 1:} Distribution of the number of the
asymmetrical
solutions depending on the chosen main domain $D$.}\\

Notice that considerable part of them (185 of the overall 2088) lie
outside of the $D=950$ area, e.g.:
 \bea
(-150, -25)(990, 945) \RA (294, 49) (546, 871) \nonumber \\
\mbox{where} \quad q=37, \quad \g_1=5, \quad \g_2=15, \quad \g_3=7,
\quad
\g_4=13,\nonumber \\
(128, 256)(990,180)  \RA (400, 200) (718, 236) \nonumber \\
\mbox{where} \quad q=20, \quad \g_1=8, \quad \g_2=15, \quad \g_3=10,
\nonumber \quad
\g_4=13,\\
(-80, -76)(980,931)  \RA (180, 171) (720, 684) \nonumber \\
\mbox{where} \quad q=761, \quad \g_1=2, \quad \g_2=7, \quad \g_3=3,
\quad \g_4=6 \nonumber \eea etc. As a whole, asymmetrical solutions
are distributed not uniformly along the wave spectrum but are rather
grouped around some specific wave numbers. For instance, the first
group of asymmetrical solutions (containing 8 solutions) appears in
the domain $D=50$, with solution
$$ (-4 , -4 )(49, 49) \RA (9, 9) (36, 36),$$
and others, while in the domains $D=60,70,80,90$ there are no new
asymmetrical solutions. The next new group (16 solutions) appears in
the domain $D=100$, and so on. From the physical point of view,
asymmetrical solutions are the most interesting ones because they
generate new wave lengths and, therefore, distribute energy through
the scales. As it was pointed out quite recently \cite{nazpok},
asymmetrical solutions play an extremely important role in wave
turbulence. Indeed, no profound understanding of turbulence can be
achieved
without studying properties which is in our agenda.\\

Numerical data for the case of planetary waves are given in the
Table 4 below:\\

\begin{tabular}{|c|c|c|c|c|c|}
\hline 
    $Domain$ & $\le 200$ & $\le 400$ & $\le 600$ & $\le 800$ & $\le 1000$\\
\hline
    $Solutions$ & 1099 & 3137 & 5664  & 8565  & 11795\\
\hline 
\end{tabular}\\

{\small Table 4. {\bf Example 2:} Distribution of the number of
solutions depending on the chosen main domain $D$.}\\

This data is presented graphically in Figs. \ref{f:2a}-\ref{f:2d}.
Number of asymmetric solutions for Example 1 (gravity waves) and
total solutions for Example 2 (planetary waves) show smooth power
growth and probably {\it are} asymptotically power functions of the
domain size $D$. On the contrary, the total solution number for
Example 1 has an unexpected twist about $D=350$ (Fig.\ref{f:2a},
shown in detail at Fig.\ref{f:2b}). This phenomenon is presently
under study.\\

 Notice that the algorithm presented here allows to find
all solutions for wave vectors belonging to the same class. For
three-wave interactions of arbitrary wave types this is always the
case.
 For $n$-wave interactions with $n>3$, however, interacting waves
 may belong to $\lfloor \frac{n}{2} \rfloor$ different classes \cite{bristol}.
 Consider for example the four-wave system
\bea \begin{cases}
    \sqrt{k_1}+\sqrt{k_2}=\sqrt{k_3}+\sqrt{k_4} \nonumber \\
    \vec{k}_1+\vec{k}_2=\vec{k}_3+\vec{k}_4 \nonumber \\
\end{cases}\eea

where $k_1$ and $k_3$ belong to one class and $k_2$ and $k_4$ - to
another one, i.e. the first equation breaks up into two independent
equations
$$ \sqrt{k_1}=\sqrt{k_3} \quad \mbox{and} \quad
\sqrt{k_2}=\sqrt{k_4}.$$ In this case, a modified form of our
generic algorithm can be applied. This will be dealt with in our
next paper.

{\bf Acknowledgement}.  E.K.  acknowledges the support of the
Austrian Science Foundation (FWF) under projects SFB F013/F1304.

\newpage

\begin{figure}[h]
\includegraphics[width=11cm,height=6cm]{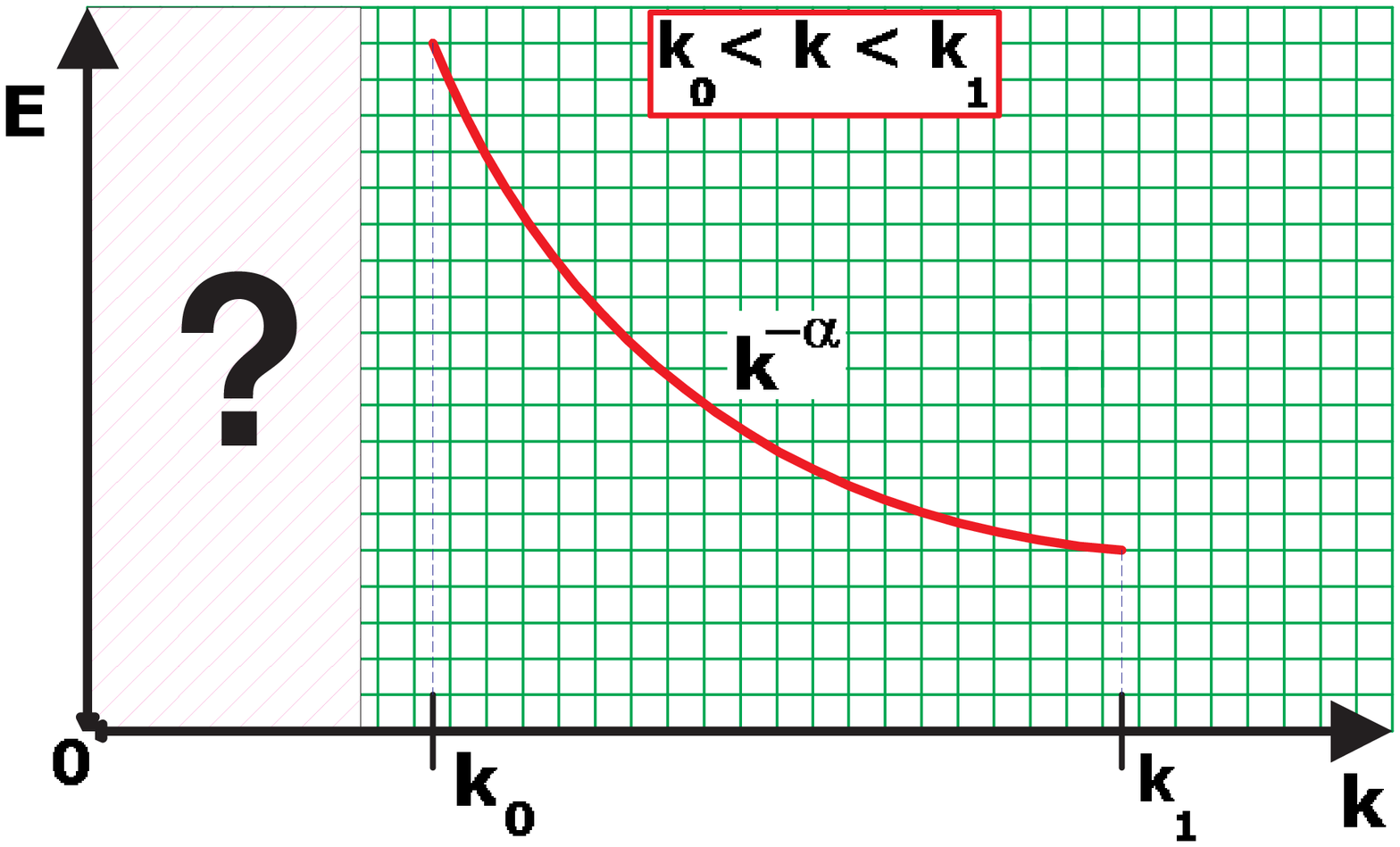}
\caption{{\bf Statistical Wave Turbulence Theory, $|\vec{k}|>k_0$:}
Dependence of the energy $E$ on the wave number $k$ is presented in
the right part of the picture; ZK-energy spectrum is shown
symbolically. Left part of the picture is not explained by
statistical wave turbulence theory.}
\end{figure}

\begin{figure}[h]
\includegraphics[width=7cm,height=4cm]{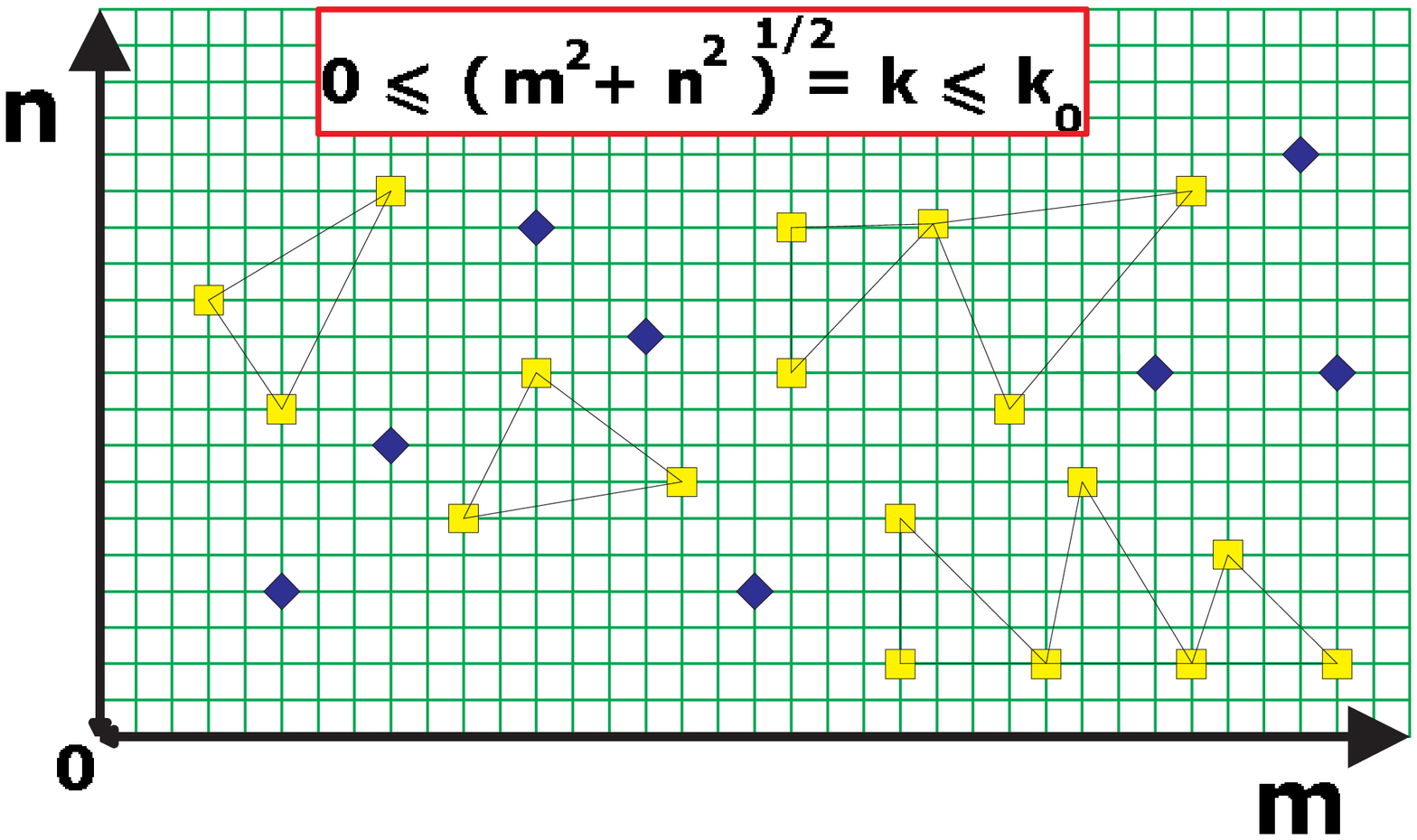}
\includegraphics[width=7cm,height=4cm]{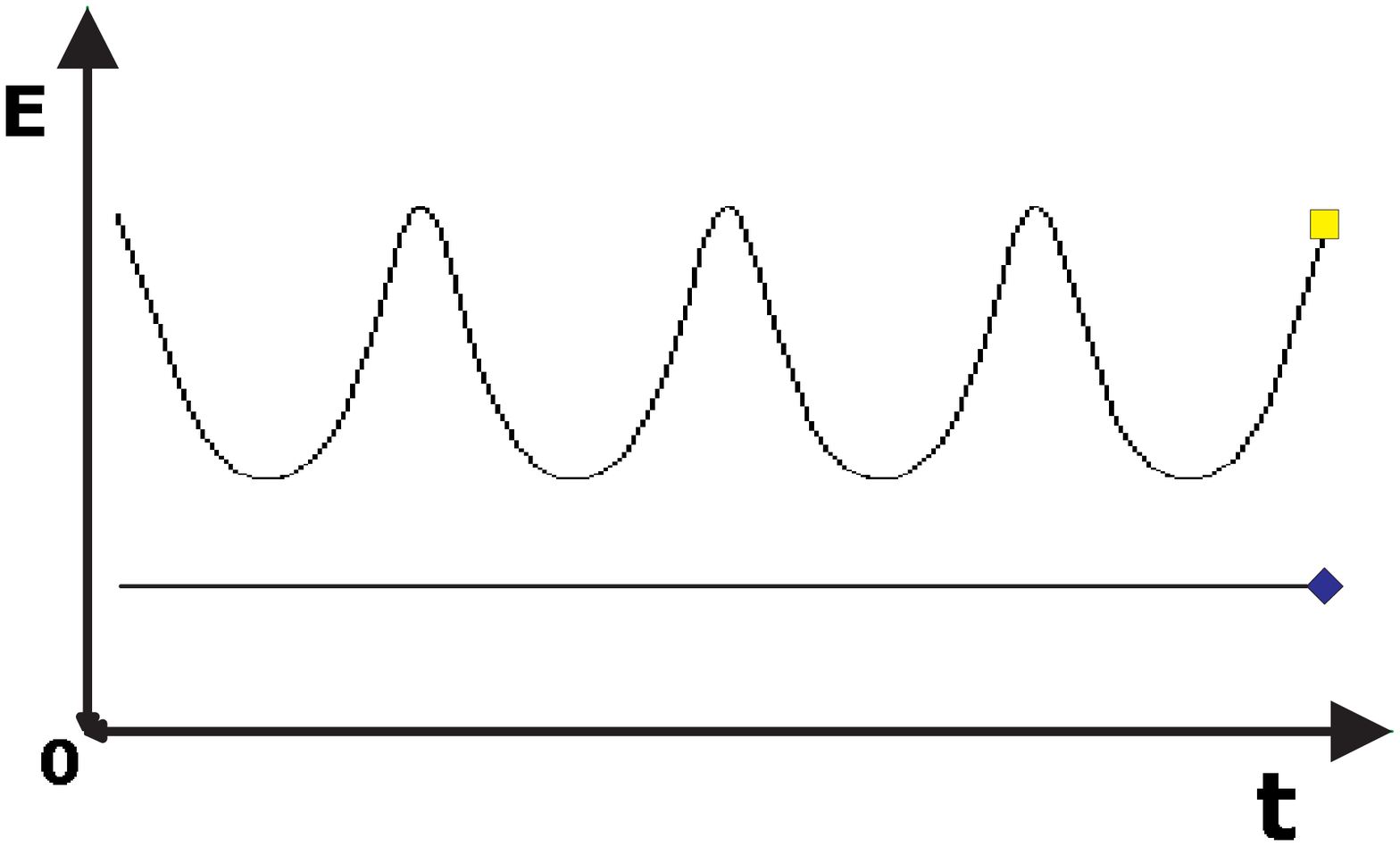}
\caption{{\bf Discrete Wave Turbulence Theory, $|\vec{k}|<k_0$:}
2D-spectral space is shown on the left panel, each node $(n,m)$ of
integer lattice corresponds to the wave vector $\vec{k}=(n,m).$
Nodes corresponding to resonantly interacting waves are depicted by
yellow squares and to non-interacting waves - by blue diamonds. On
the right panel, dependence of the energy $E$ on time $t$ for both
types of waves is shown}
\end{figure}

\begin{figure}[h]
\includegraphics[width=11cm,height=6cm]{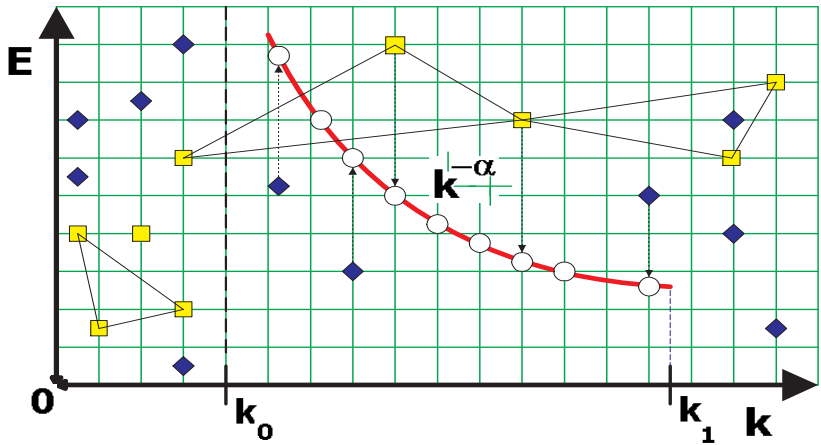}
\caption{{\bf Laminated Wave Turbulence Theory, arbitrary
$|\vec{k}|$:} Discrete and statistical layers of turbulence co-exist
in many wave systems. ZK-energy spectrum contains "holes" in the
nodes of the integer lattice which are depicted by empty circles.}
\end{figure}

\newpage
\begin{figure}\begin{center}
\vskip -0.5cm  \hskip -0.5cm
 \includegraphics[width=8.5cm,height=6.5cm]{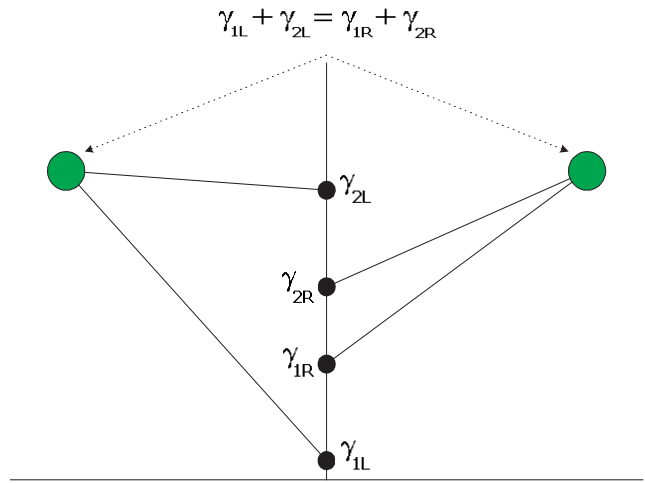}
 \end{center}
\caption{\label{f:1a} Weight diagram, Case 1: no degeneration}
\end{figure}

\newpage
\begin{figure}\begin{center}
\vskip -0.5cm  \hskip -0.5cm
 \includegraphics[width=8.5cm,height=6.5cm]{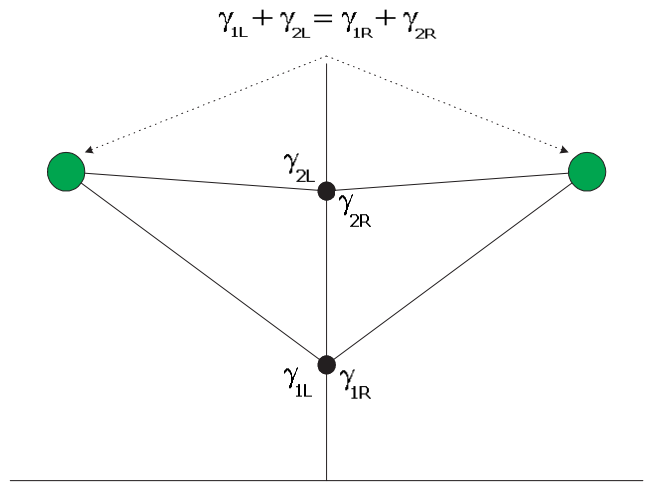}
 \end{center}
\caption{\label{f:1b} Weight diagram, Case 2: left-right
degeneration}
\end{figure}

\newpage
\begin{figure}\begin{center}
\vskip -0.5cm  \hskip -0.5cm
 \includegraphics[width=8.5cm,height=6.5cm]{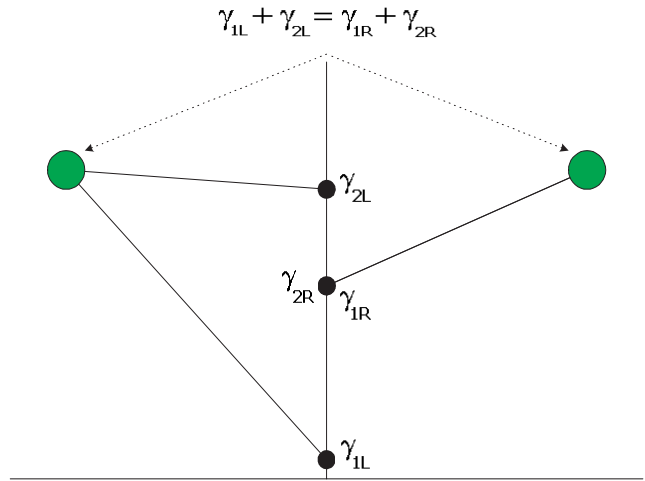}
 \end{center}
\caption{\label{f:1c} Weight diagram, Case 3: right degeneration}
\end{figure}
\newpage
\begin{figure}\begin{center}
\vskip -0.5cm  \hskip -0.5cm

\newpage
 \includegraphics[width=8.5cm,height=6.5cm]{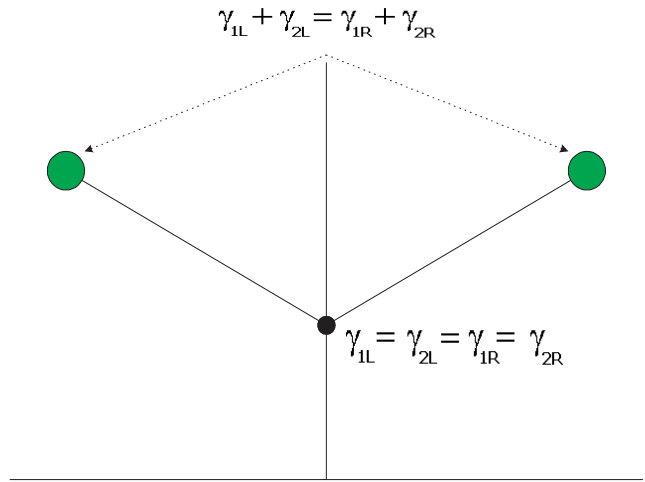}
 \end{center}
\caption{\label{f:1d} Weight diagram, Case 4: total degeneration}
\end{figure}
\newpage
\begin{figure}\begin{center}
\vskip -0.5cm  \hskip -0.5cm

\newpage
 \includegraphics[width=10cm,height=6.5cm]{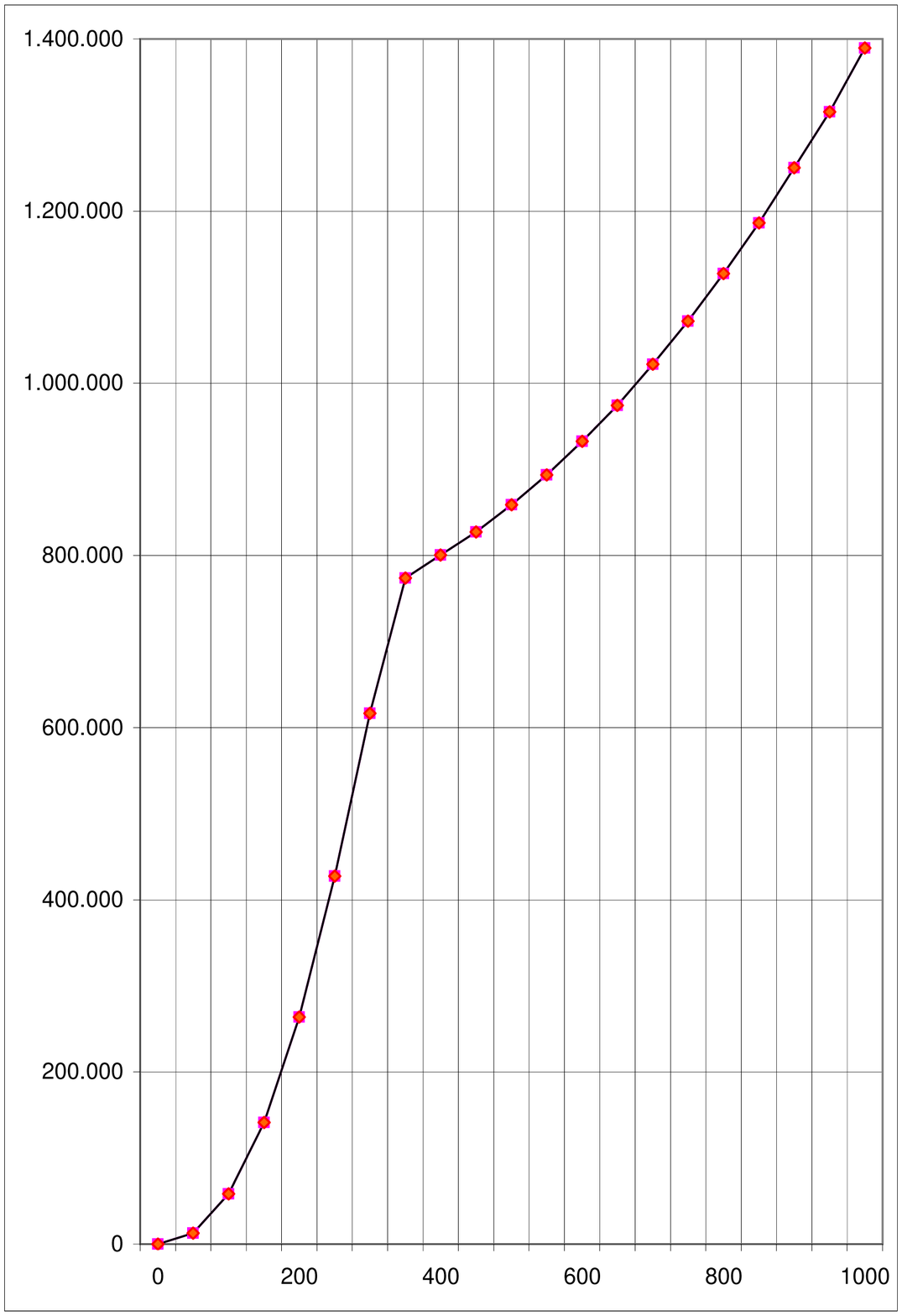}
 \end{center}
\caption{\label{f:2a} Number of solutions in partial domains}
\end{figure}
\newpage
\begin{figure}\begin{center}
\vskip -0.5cm  \hskip -0.5cm

\newpage
 \includegraphics[width=10cm,height=6.5cm]{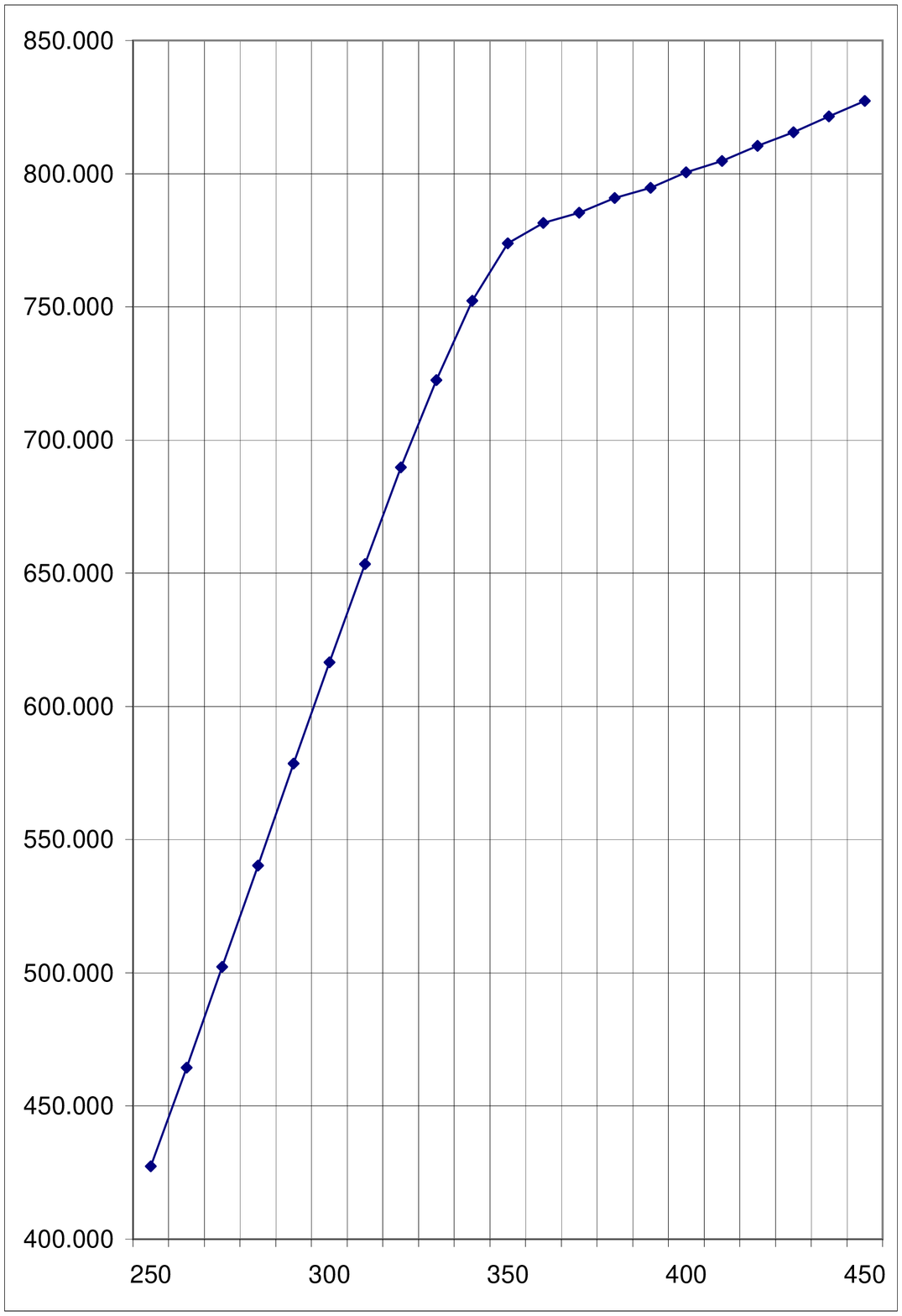}
 \end{center}
\caption{\label{f:2b} Twist range of Fig. \ref{f:2a} zoomed}
\end{figure}

\newpage
\begin{figure}\begin{center}
\vskip -0.5cm  \hskip -0.5cm
 \includegraphics[width=10cm,height=6.5cm]{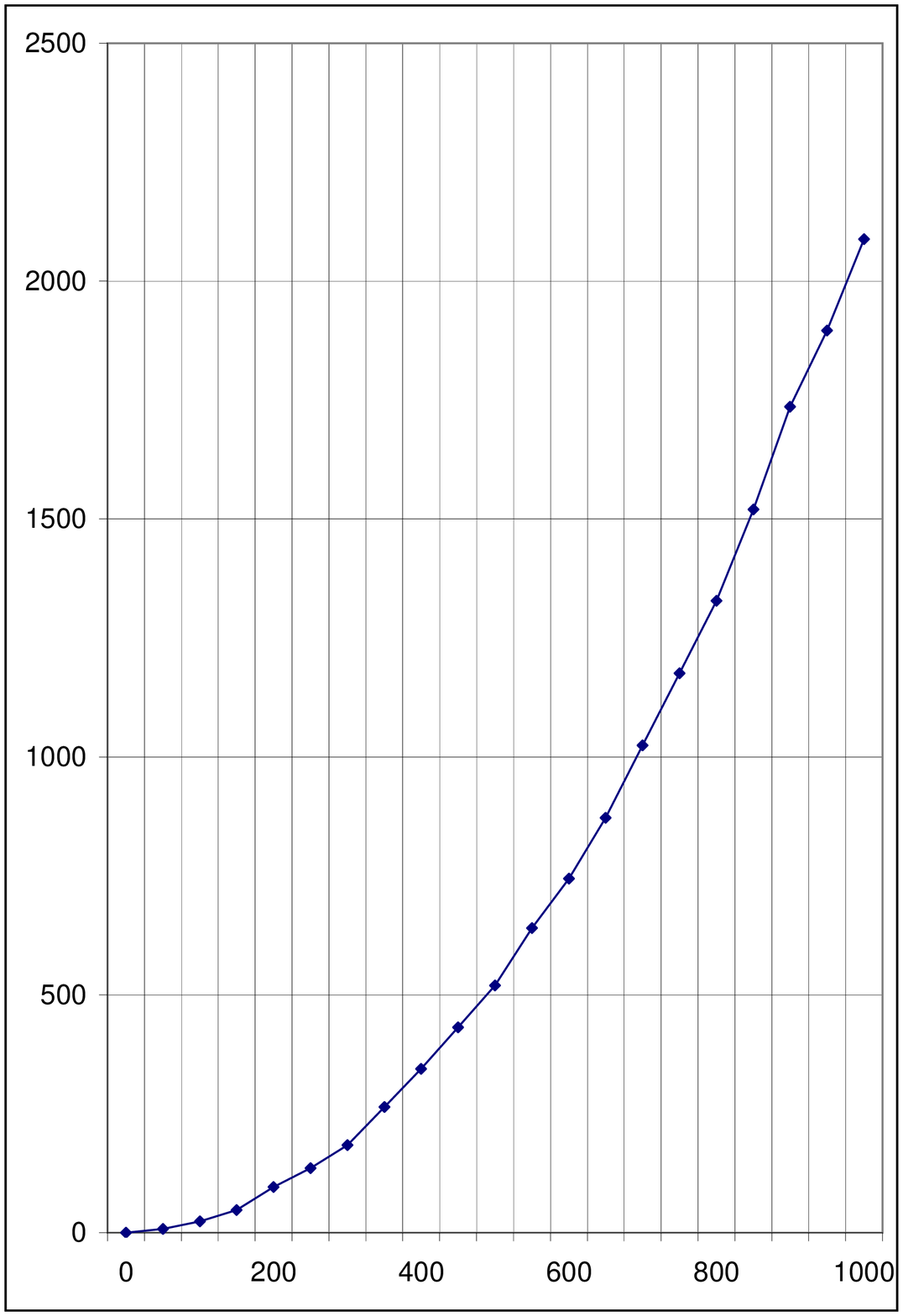}
 \end{center}
\caption{\label{f:2c} Number of asymmetric solutions in partial
domains}
\end{figure}

\newpage
\begin{figure}\begin{center}
\vskip -0.5cm  \hskip -0.5cm
 \includegraphics[width=10cm,height=6.5cm]{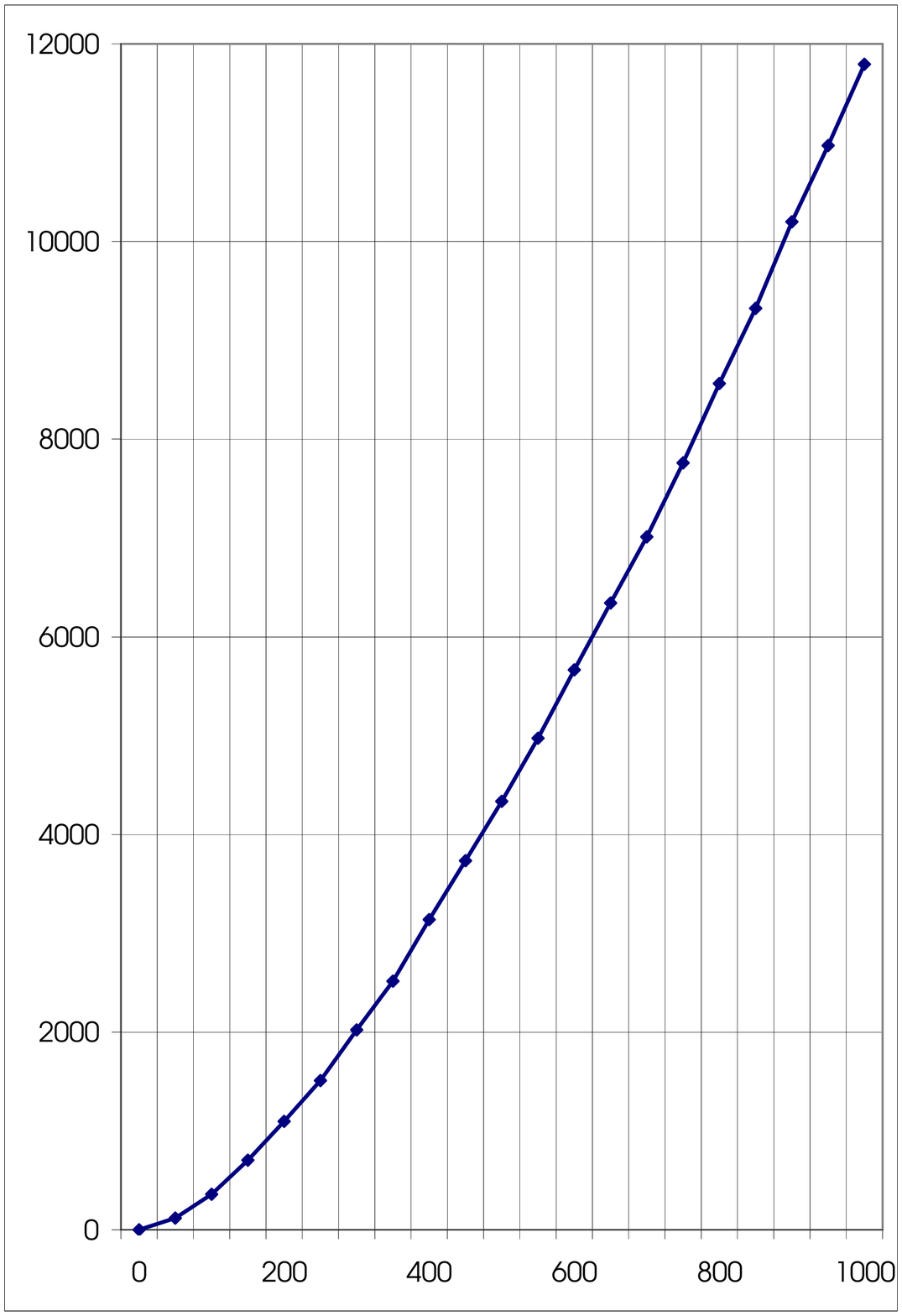}
 \end{center}
\caption{\label{f:2d} Example 2: Number of solutions in partial
domains}
\end{figure}

\end{document}